\documentclass[manuscript, screen, oneside]{acmart}
\AtBeginDocument{%
  }

\setcopyright{acmlicensed}
\copyrightyear{2026}
\acmYear{2026}
\acmConference[NLPAI '2026]{2026 7th International Conference on Natural Language Processing and Artificial Intelligence}{July 10--13, 2026}{Urumqi, China}
\acmISBN{979-8-4007-2497-8}

\usepackage{booktabs}
\usepackage{multirow}
\begin{document}
\pagestyle{plain}
\title{Coarse-to-Fine Diffusion with Residual Prior for Low-Dose CT Denoising}

\author{Ping Wu}
\affiliation{%
  \institution{University of Electronic Science and Technology of China}
  \city{Chengdu}
  \country{China}
}
\email{pingwu0616@163.com}
\orcid{0009-0005-0932-0451}

\author{Pengcheng Zheng}
\affiliation{%
  \institution{University of Electronic Science and Technology of China}
  \city{Chengdu}
  \country{China}}
\email{zpc777@std.uestc.edu.cn}
\orcid{0009-0004-3323-9170}

\author{Jiahao Zhang}
\affiliation{%
 \institution{University of Electronic Science and Technology of China}
 \city{Chengdu}
 \country{China}}
\email{zhangjiahao010706@163.com}
\orcid{0009-0001-1930-3621}

\author{Bohao Chen}
\affiliation{%
  \institution{University of Electronic Science and Technology of China}
  \city{Chengdu}
  \country{China}}
\email{chenbohao12138@gmail.com}
\orcid{0009-0002-7153-8185}

\author{Yazhou Ren}
\affiliation{%
  \institution{University of Electronic Science and Technology of China}
  \city{Chengdu}
  \country{China}}
\affiliation{%
  \institution{Shenzhen Institute for Advanced Study, University of Electronic Science and Technology of China
}
  \city{Shenzhen}
  \country{China}
}
\email{yazhou.ren@uestc.edu.cn}
\orcid{0000-0001-7705-4603}

\author{Xiaorong Pu}
\authornote{Corresponding author.}
\affiliation{%
  \institution{University of Electronic Science and Technology of China}
  \city{Chengdu}
  \country{China}}
\affiliation{%
  \institution{Shenzhen Institute for Advanced Study, University of Electronic Science and Technology of China
}
  \city{Shenzhen}
  \country{China}
}
\email{puxiaor@uestc.edu.cn}
\orcid{0000-0001-7387-7194}

\renewcommand{\shortauthors}{Wu et al.}

\begin{abstract}
  Low-dose CT (LDCT) reduces radiation exposure but often introduces severe noise and streak artifacts that obscure subtle anatomical structures. Existing diffusion-based denoising methods commonly rely on implicit assumptions about the noise distribution, which may not hold in practical LDCT scenarios and can degrade denoising performance. To address this limitation, we propose a Residual-Guided Diffusion framework that effectively incorporates an explicit difference prior to inform the diffusion process of noise distribution. The difference prior is computed as the residual between the LDCT image and a coarse clean estimate, constraining diffusion sampling to a plausible residual space. We further adopt a coarse-to-fine strategy in which a pre-trained RED-CNN provides the coarse estimate to anchor global anatomy, and the diffusion model refines the remaining details under the prior guidance. For efficiency, we employ accelerated sampling with 20 steps at inference. Experiments on three representative datasets demonstrate that our approach outperforms existing methods in both quantitative and qualitative performance measures.
\end{abstract}




\maketitle

\section{Introduction}
Computed Tomography (CT) is a cornerstone of modern medical diagnosis. However, radiation exposure in normal-dose
CT (NDCT) scans may cause unavoidable health damage
and even increase the risk of cancers~\cite{smith2025projected}. Lowering the dose improves patient safety, but LDCT acquisition often suffers from severe noise and streak artifacts due to photon starvation~\cite{manduca2009projection}. These degradations reduce image quality and can weaken diagnostic confidence. The impact is most pronounced for subtle high frequency findings such as small pulmonary nodules and early stage hepatic lesions. Recovering NDCT quality images from noisy LDCT measurements therefore remains a challenging and clinically important task.

Deep learning (DL) has become the dominant paradigm for LDCT denoising and has received broad attention in medical imaging. Early Convolutional Neural Networks (CNNs)-based methods~\cite{chen2017low}, such as RED-CNN~\cite{chen2017redcnn} and WGAN-VGG~\cite{yang2018wgan}, achieved strong results by learning a direct mapping from LDCT to NDCT. Subsequent advances, including MAP-NN~\cite{shan2019mapnn} and the Transformer-based CTformer~\cite{wang2023ctformer}, further exploit larger receptive fields and global context to suppress noise and artifacts. Despite this progress, many approaches are trained mainly with pixel level fidelity losses, such as MSE or $L_{1}$ loss. These objectives tend to produce estimates close to the conditional mean of the posterior distribution. This bias often leads to over-smoothing and attenuates fine textures and subtle structures~\cite{blau2018perception}, which may reduce clinical utility in downstream diagnostic assessment and reporting.

Recently, diffusion-based methods have been explored for LDCT denoising due to their success in image generation~\cite{liu2025dndp,gao2025need,kim2025dose}. Denoising Diffusion Probabilistic Models (DDPMs)~\cite{ho2020ddpm} provide a generative framework and often achieve better texture fidelity than GANs-based methods~\cite{goodfellow2014generative} in many settings. Recent studies have demonstrated that incorporating reliable priors can enhance the performance of medical image restoration tasks~\cite{zheng2025lightweight,du2024structure}. In LDCT denoising, the LDCT image is commonly used as a condition to guide the reverse diffusion process toward a clean target. However, applying standard conditional diffusion to LDCT remains difficult. The reverse process relies on stochastic sampling in a high dimensional space. When noise and streak artifacts are pronounced, conditioning on the LDCT input alone may be insufficient. This can lead to structural inconsistency or hallucinated patterns that appear realistic but are anatomically implausible~\cite{pfaff2024no}. The risk increases when the noise component overwhelms the anatomical signal. These observations motivate a key question: can an explicit noise prior constrain diffusion sampling and improve anatomical fidelity in LDCT denoising?

To address this issue, we propose a Residual-Guided Diffusion framework with an explicit difference prior for LDCT denoising. Concretely, we employ a pre-trained RED-CNN to produce a coarse anatomical estimate, and compute the residual between the LDCT input and this estimate as a difference prior. The prior is concatenated with the LDCT condition and jointly encoded to guide the diffusion model. This design provides constraints beyond the LDCT observation alone. The difference prior captures spatially varying degradations such as streak artifacts. It also retains high frequency components that are often attenuated by deterministic CNN denoisers. Conditioning diffusion refinement on this prior promotes residual-consistent reconstruction. It improves anatomical consistency and preserves subtle diagnostically relevant structures. Our main contributions are summarized as follows:
\begin{itemize}
    \item We reformulate the LDCT denoising task as a  difference prior-constrained inference problem, which significantly mitigates the ill-posedness of diffusion-based denoising under noisy observations.
    
    \item We propose a residual-guided diffusion inference strategy that constrains the stochastic sampling process in the residual (noise-dominated) space, leading to substantially improved anatomical consistency compared with unconstrained image-space diffusion models.
    
    \item We develop an efficient coarse-to-fine framework with accelerated sampling, achieving competitive quantitative and qualitative performance on three datasets.
\end{itemize}

\begin{figure}[t]
  \includegraphics[width=\columnwidth]{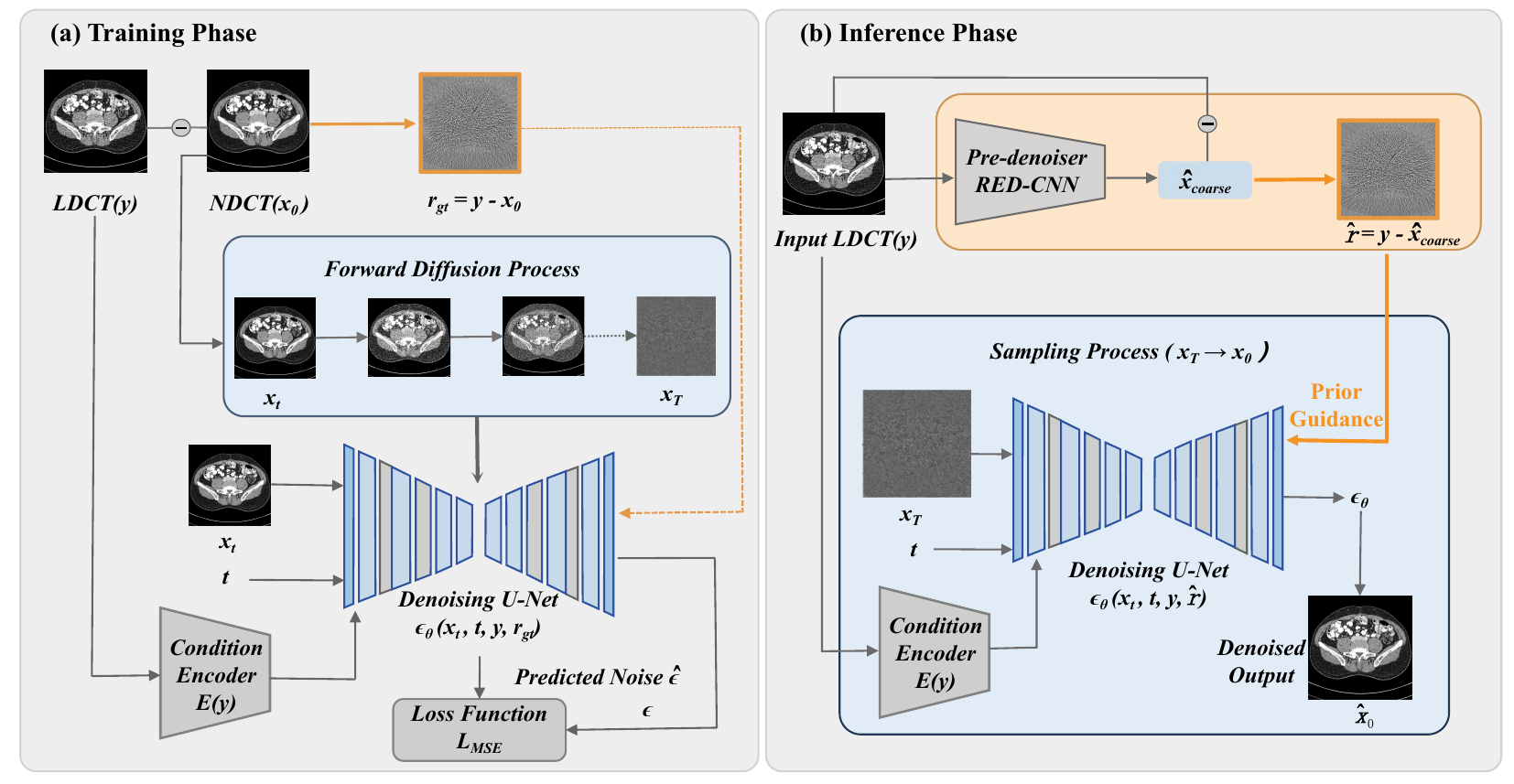}
  \caption{Workflow of the proposed Residual-Guided Diffusion model. A coarse estimate obtained by RED-CNN is used to construct an explicit difference prior map $r$, which is incorporated as a conditioning signal to guide the subsequent diffusion-based refinement of the noisy LDCT input $y$.}
  \Description{The overall workflow of the proposed Residual-Guided Diffusion model, consisting of a training phase and an inference phase.}
  \label{fig:framework}
\end{figure}

\section{Method}
We propose a Residual-Guided Diffusion framework for LDCT denoising. The framework combines fast CNN-based  restoration with diffusion-based refinement for fine-detail recovery. Fig.~\ref{fig:framework} illustrates our proposed two stage coarse-to-fine pipeline:

1) \textbf{Prior generation stage}: A pre-trained RED-CNN is applied to the LDCT input to produce a coarse estimate of the corresponding NDCT image. We then compute an explicit difference prior as the residual between the LDCT observation and the coarse estimate.

2) \textbf{Diffusion refinement stage}: The estimated difference prior is incorporated as an additional conditional input to the diffusion model. The model performs residual-guided refinement and progressively reconstructs a high-quality NDCT image. The final output is the refined NDCT reconstruction.

\subsection{Problem Formulation}

Let $x \in \mathbb{R}^{H \times W \times C}$ denote a high-quality NDCT image and let $y \in \mathbb{R}^{H \times W \times C}$ denote the corresponding LDCT observation. The LDCT degradation process can be written as
\begin{equation}
y = x + n,
\label{eq:degradation}
\end{equation}
where $n$ represents the mixed Poisson--Gaussian noise induced by photon starvation. This type of noise is non-stationary and spatially signal-dependent in practice~\cite{lee2018poisson}.

The goal of LDCT denoising is to recover $x$ from $y$. This problem is often formulated as a MAP estimation:
\begin{equation}
\hat{x} = \arg \max_x \log p(y|x) + \log p(x).
\label{eq:map}
\end{equation}
Here, $p(y|x)$ denotes the likelihood implied by the degradation model, and $p(x)$ encodes the prior distribution of clean NDCT images. Maximizing Eq.~\ref{eq:map} is equivalent to maximizing $\log p(x|y)$ up to an additive constant $\log p(y)$. Instead of directly optimizing the MAP objective, diffusion models perform iterative denoising to approximate sampling from the conditional distribution $p(x|y)$~\cite{khader2023denoising}.

\subsection{Difference Prior Construction}

Balancing noise suppression and texture preservation remains a key challenge in LDCT denoising. Deterministic CNN-based approaches are typically trained with pixel-wise fidelity losses such as MSE. Although such objectives often yield stable anatomical restoration, they tend to bias predictions toward over-smoothed solutions, thereby attenuating high-frequency diagnostically relevant details. However, when LDCT measurements contain substantial noise and streak artifacts, diffusion sampling conditioned only on the LDCT input can be insufficiently constrained. This may lead to structural inconsistency or hallucinated patterns. It can also produce non-existent structures or misinterpret streak artifacts as biological tissue.

To improve both structural fidelity and detail recovery, we adopt a coarse-to-fine strategy and construct an explicit difference prior. Specifically, we employ a pre-trained RED-CNN as a coarse estimator $F_{\mathrm{cnn}}$ to capture dominant anatomical structures:
\begin{equation}
\hat{x}_{\text{coarse}} = F_{\mathrm{cnn}}(y).
\label{eq:coarse}
\end{equation}
Here, $\hat{x}_{\text{coarse}}$ denotes the coarse estimate of $x$. While this estimate typically has higher apparent SNR, it does not explicitly describe the spatial distribution of noise. We therefore define $r$ as the residual between the LDCT observation and the coarse estimate:

\begin{equation}
r = y - \hat{x}_{\text{coarse}}. 
\end{equation}
The residual map $r$ provides an explicit representation of noise-dominated in the LDCT image. We incorporate $r$ as an additional conditional input to the diffusion model to guide fine-grained reconstruction. This design encourages residual-consistent refinement instead of unconstrained generation in the full image space. It reduces sampling ambiguity and improves anatomical consistency. 

During the training phase, we use the ground-truth residual $r_{\mathrm{gt}} = y - x$ as the conditioning prior.
During the inference phase, we replace it with the estimated residual $\hat{r} = y - \hat{x}_{\mathrm{coarse}}$. While this substitution introduces a potential train-test discrepancy in the prior distribution, empirical observations show that our model remains highly robust to this mismatch. The deviation between the estimated residual and the ground truth introduces only negligible performance drops. This stability arises because the diffusion model is trained to learn the general distribution of residual representations rather than over-fitting to exact pixel values, allowing it to correctly recover key anatomical structures even under an estimated prior.

\subsection{Residual-Guided Diffusion Process}

We leverage a conditional diffusion probabilistic model to refine texture details~\cite{saharia2022sr3}. In the forward process, Gaussian noise is gradually added to the data distribution. In the reverse process, the clean image $x_0$ is recovered from pure noise $x_T$, conditioned on the LDCT input $y$ and the difference prior map $r$. The reverse process is modeled as a Markov chain~\cite{ho2020ddpm} with learned Gaussian transitions:
\begin{equation}
p_\theta(x_{t-1} | x_t, y, r)
= \mathcal{N}\!\left(x_{t-1}; \mu_\theta(x_t, t, y, r), \sigma_t^2 \mathbf{I}\right), 
\label{eq:reverse}
\end{equation}
where $t \in \{T, \dots, 1\}$ denotes the timestep index. In Eq.~\ref{eq:reverse}, $\mathcal{N}(\cdot; \mu, \Sigma)$ represents a Gaussian distribution with mean $\mu$ and covariance $\Sigma$, and $\mathbf{I}$ denotes the identity matrix. The variance schedule $\sigma^2_t$ is predefined following standard DDPM settings, while the mean $\mu_\theta$ is the core component to be learned. Specifically, $\mu_\theta$ is implicitly parameterized by a deep neural network $\epsilon_\theta$ with learnable parameters $\theta$ optimized during training.

As formulated in Eq.~\ref{eq:concat}, this network takes the channel-wise concatenation of the noisy state $x_t$, the LDCT condition $y$, and the difference prior $r$ as input:
\begin{equation}
\text{Input} = \mathrm{Concat}(x_t, y, r).
\label{eq:concat}
\end{equation}
This formulation differs from standard conditional diffusion,
which conditions only on the LDCT image. By explicitly injecting the difference prior map $r$, the network is relieved from learning global anatomy and can focus on distinguishing streaking artifacts from subtle pathological textures within the residual domain. The network is trained to predict the added noise $\epsilon$ by minimizing the following objective:
\begin{equation}
\mathcal{L}
= \mathbb{E}_{x_0,\epsilon,t,y}
\left[
\left\|
\epsilon - \epsilon_\theta(x_t, t, y, r)
\right\|_2^2
\right].
\label{eq:loss}
\end{equation}
During the inference phase, the model iteratively samples from $x_T$ to $x_0$. With the guidance of the difference prior map $r$, the stochastic search space of the diffusion model is constrained, reducing the risk of non-existent structures while faithfully restoring low-contrast anatomical details.

\begin{table}[t]
  \centering
  \small
  \caption{Quantitative comparisons on three datasets: Mayo 2016, Mayo 2020, and LoDoPaB-CT. The best and second-best results are highlighted and underlined respectively.}
  \label{tab:quantitative_comparison}
  \begin{tabular}{lcccccc}
    \toprule
    \multirow{2}{*}{Method} & \multicolumn{2}{c}{Mayo 2016} & \multicolumn{2}{c}{Mayo 2020} & \multicolumn{2}{c}{LoDoPaB-CT} \\
    \cmidrule(lr){2-3} \cmidrule(lr){4-5} \cmidrule(lr){6-7}
    & PSNR$\uparrow$ & SSIM$\uparrow$ & PSNR$\uparrow$ & SSIM$\uparrow$ & PSNR$\uparrow$ & SSIM$\uparrow$ \\
    \midrule
    RED-CNN   & 41.58 & 0.9682 & 33.42 & 0.7670 & 32.37 & 0.7996 \\
    WGAN-VGG  & 41.71 & 0.9625 & 32.23 & 0.7079 & 35.84 & 0.8265 \\
    Restormer & 41.99 & 0.9699 & \underline{35.89} & \underline{0.7684} & 36.12 & 0.8427 \\
    CTformer  & 41.92 & 0.9601 & 33.74 & 0.7396 & 34.72 & 0.8190 \\
    Fast-DDPM & 42.03 & 0.9641 & 32.73 & 0.7379 & 34.86 & 0.8214 \\
    CoreDiff  & 42.58 & 0.9643 & 33.39 & 0.7612 & 35.19 & 0.8277 \\
    FoundDiff & \underline{42.62} & \underline{0.9722} & 34.51 & 0.7540 & \underline{36.46} & \underline{0.8361} \\
    \midrule
    \textbf{Ours} & \textbf{43.27} & \textbf{0.9787} & \textbf{36.44} & \textbf{0.7755} & \textbf{36.93} & \textbf{0.8512} \\
    \bottomrule
  \end{tabular}
\end{table}

\section{Experiment}
\subsection{Datasets}

We evaluate our method on three representative datasets, namely the Mayo 2016 dataset \cite{mccollough2017mayo2016}, the Mayo 2020 dataset \cite{moen2021mayo2020}, and the LoDoPaB-CT dataset \cite{leuschner2021lodopab}. The Mayo 2016 dataset provides paired NDCT and LDCT images from 10 patients. We use 8 patients with 4,553 slices for training, 1 patient with 823 slices for validation, and 1 patient with 560 slices for testing. The Mayo 2020 dataset contains LDCT images acquired at diverse and partially unspecified dose levels. We perform zero-shot testing on 713 slices from 2 patients. Specifically, all methods are trained only on the Mayo 2016 training set and directly tested on Mayo 2020, without using any Mayo 2020 images for training, validation, fine-tuning, hyperparameter tuning, or model selection. For the LoDoPaB-CT dataset, we report results on the first 100 images of the official test set to reduce computational cost while maintaining comparability.

\subsection{Implementation Details}

Our framework is implemented using PyTorch 2.5.1 and trained on a server with four NVIDIA RTX 4090 GPUs. We adopt a two-stage training strategy. First, the RED-CNN estimator is pre-trained until convergence. During the subsequent training of the diffusion model, we freeze the parameters of RED-CNN to maintain a consistent difference prior. The diffusion model is trained for $100,000$ iterations with a batch size of $8$ and a patch size of $256 \times 256$. We use the Adam optimizer~\cite{kingma2014adam} with a  learning rate of $1 \times 10^{-4}$ and cosine learning rate scheduling. During the inference phase, we employ an accelerated sampling strategy with 20 steps to balance reconstruction quality and computational efficiency.

\begin{figure}[t]
  \centering
  \includegraphics[width=\columnwidth]{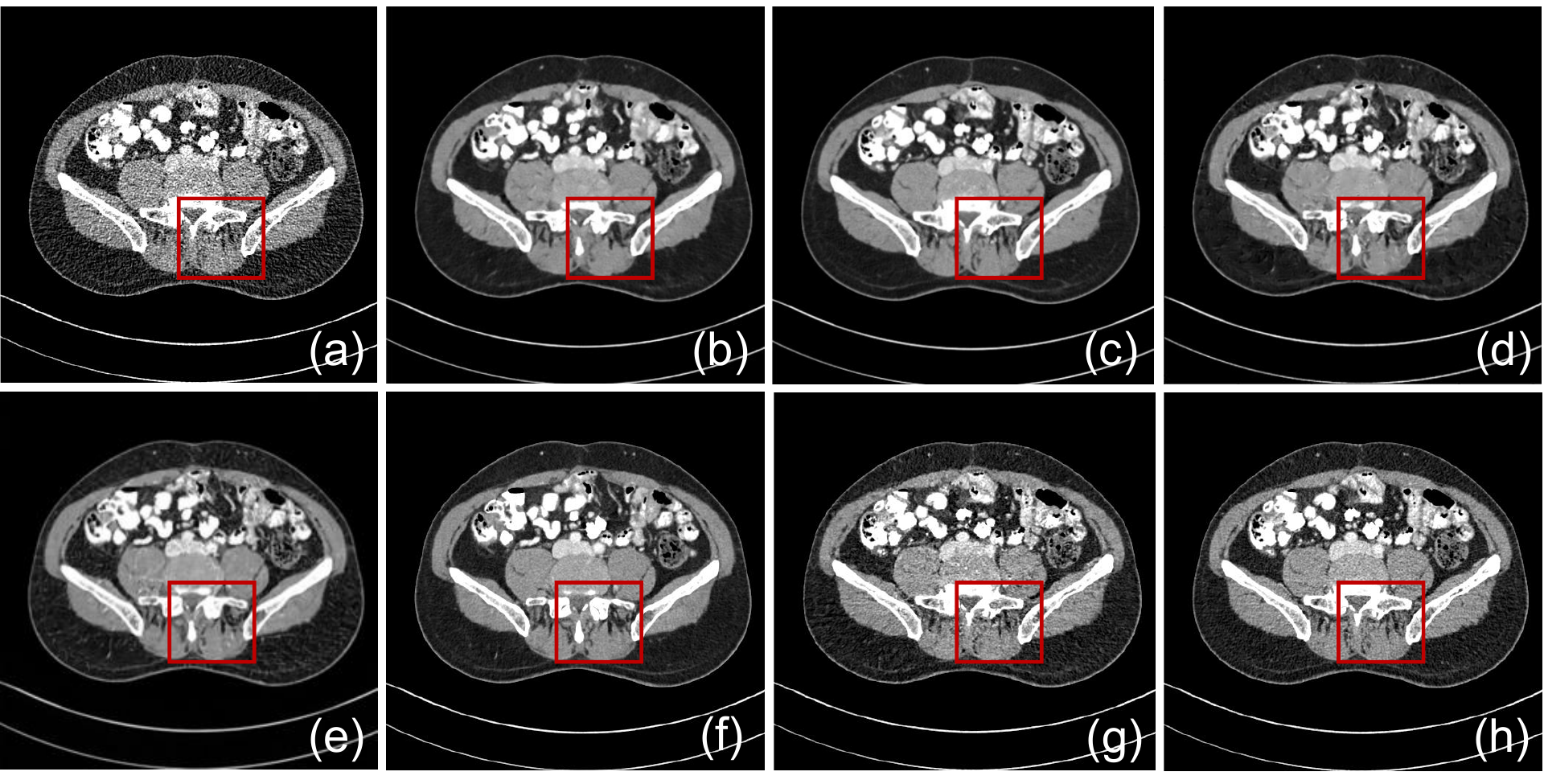}
  \caption{The denoised results of different methods on the Mayo 2016 dataset. (a) LDCT, (b) RED-CNN, (c) Restormer, (d) CTformer, (e) Fast-DDPM,  (f) CoreDiff, (g) Ours and (h) NDCT. The display window is [-160, 240] HU.}
  \Description{The visual results.}
\label{fig:visual results}
\end{figure}

\begin{figure}[t]
  \centering
  \includegraphics[width=\columnwidth]{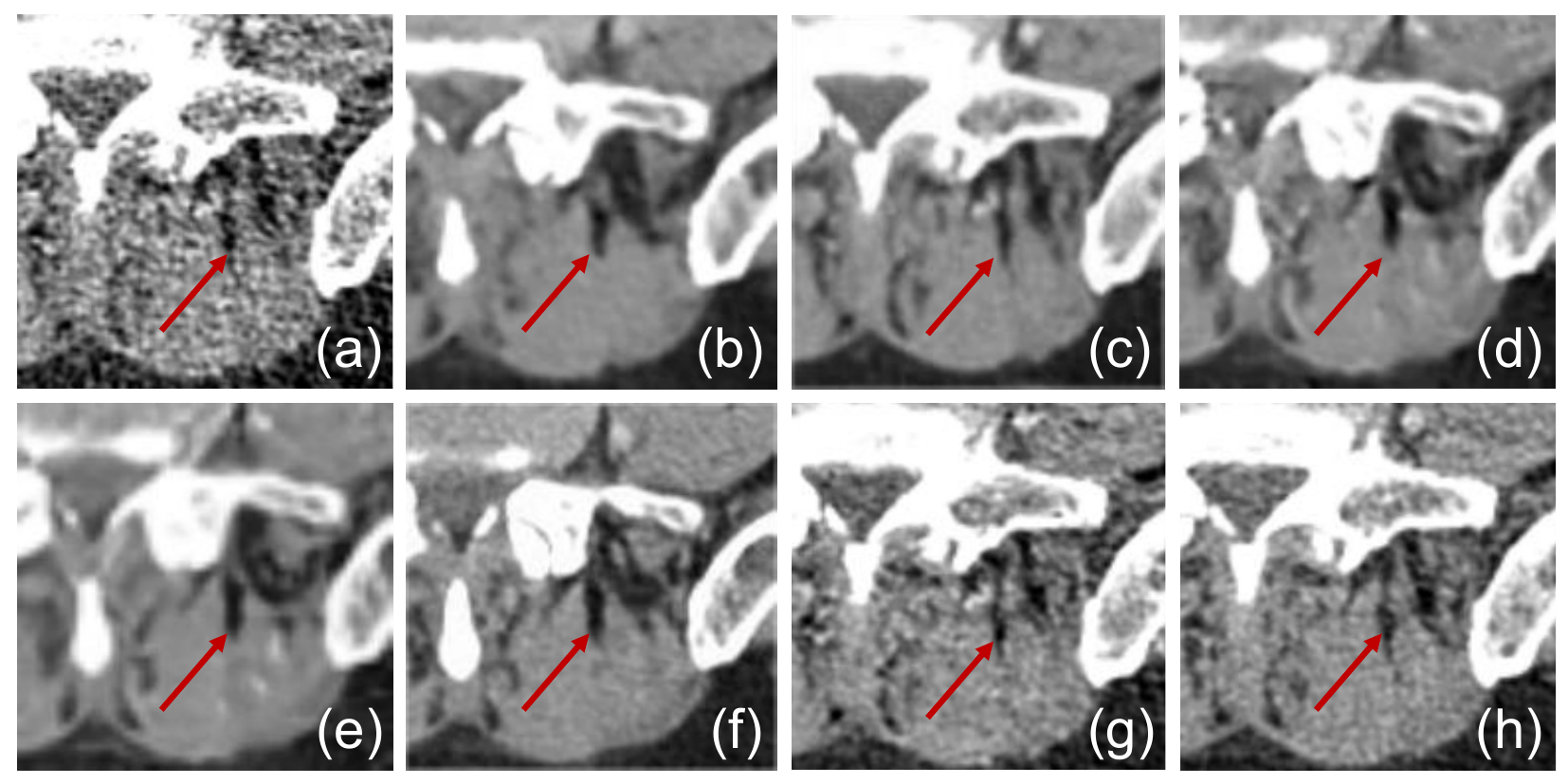}
  \caption{The ROIs of the rectangle marked in Figure~\ref{fig:visual results}. (a) LDCT, (b) RED-CNN, (c) Restormer, (d) CTformer, (e) Fast-DDPM, (f) CoreDiff, (g) Ours and (h) NDCT.}
  \Description{The ROT results.}
\label{fig:ROI results}
\end{figure}

\subsection{Visual Comparisons}

Fig.~\ref{fig:visual results} presents a visual comparison of the denoised results on the $1$ mm thin-slice test samples. We compare our method against seven representative baselines: RED-CNN~\cite{chen2017redcnn}, WGAN-VGG~\cite{yang2018wgan}, Restormer \cite{zamir2022restormer}, CTformer~\cite{wang2023ctformer}, Fast-DDPM~\cite{jiang2025fast}, CoreDiff~\cite{gao2023corediff} and FoundDiff~\cite{chen2025founddiff}. The denoising performance of each method is quantitatively evaluated using PSNR and SSIM metrics. Table~\ref{tab:quantitative_comparison} summarizes the average PSNR and SSIM results of all methods.

As shown in the zoomed-in ROIs in Fig.~\ref{fig:ROI results}, the LDCT input is severely affected by photon starvation noise and streak artifacts. RED-CNN effectively suppresses noise but tends to over-smooth fine structures.  Restormer and CTformer preserve global context better, but may still blur low-contrast details or leave streaks. Diffusion baselines such as Fast-DDPM, CoreDiff and FoundDiff recover high-frequency details. However, stochastic sampling can still introduce structural inconsistency and spurious textures.

Our method produces results that are visually closest to the NDCT ground truth. It achieves a better balance between noise removal and texture preservation. Conditioning on the difference prior encourages refinement in the residual domain and improves anatomical consistency. As indicated by the red arrows, our method better preserves small low-contrast structures and suppresses streak artifacts without introducing spurious textures.

\begin{table}[t]
  \centering
  \caption{Ablation study of the difference prior and sampling steps. We compare the performance with and without the difference prior under different sampling steps. The best results are highlighted.}
  \label{tab:ablation}
  \begin{tabular}{lccc}
    \toprule
    Sampling Steps & Difference Prior & PSNR$\uparrow$ & SSIM$\uparrow$ \\
    \midrule
    10 steps  & w/ Prior  & 42.16 & 0.9652 \\
    20 steps  & w/o Prior & 42.31 & 0.9670 \\
    20 steps  & w/ Prior  & \textbf{43.27} & \textbf{0.9787} \\
    50 steps  & w/ Prior  & 41.74 & 0.9582 \\
    100 steps & w/ Prior  & 41.12 & 0.9563 \\
    500 steps & w/ Prior  & 41.05 & 0.9498 \\
    \bottomrule
  \end{tabular}
\end{table}

\begin{figure}[t]
  \centering
  \includegraphics[width=1\linewidth]{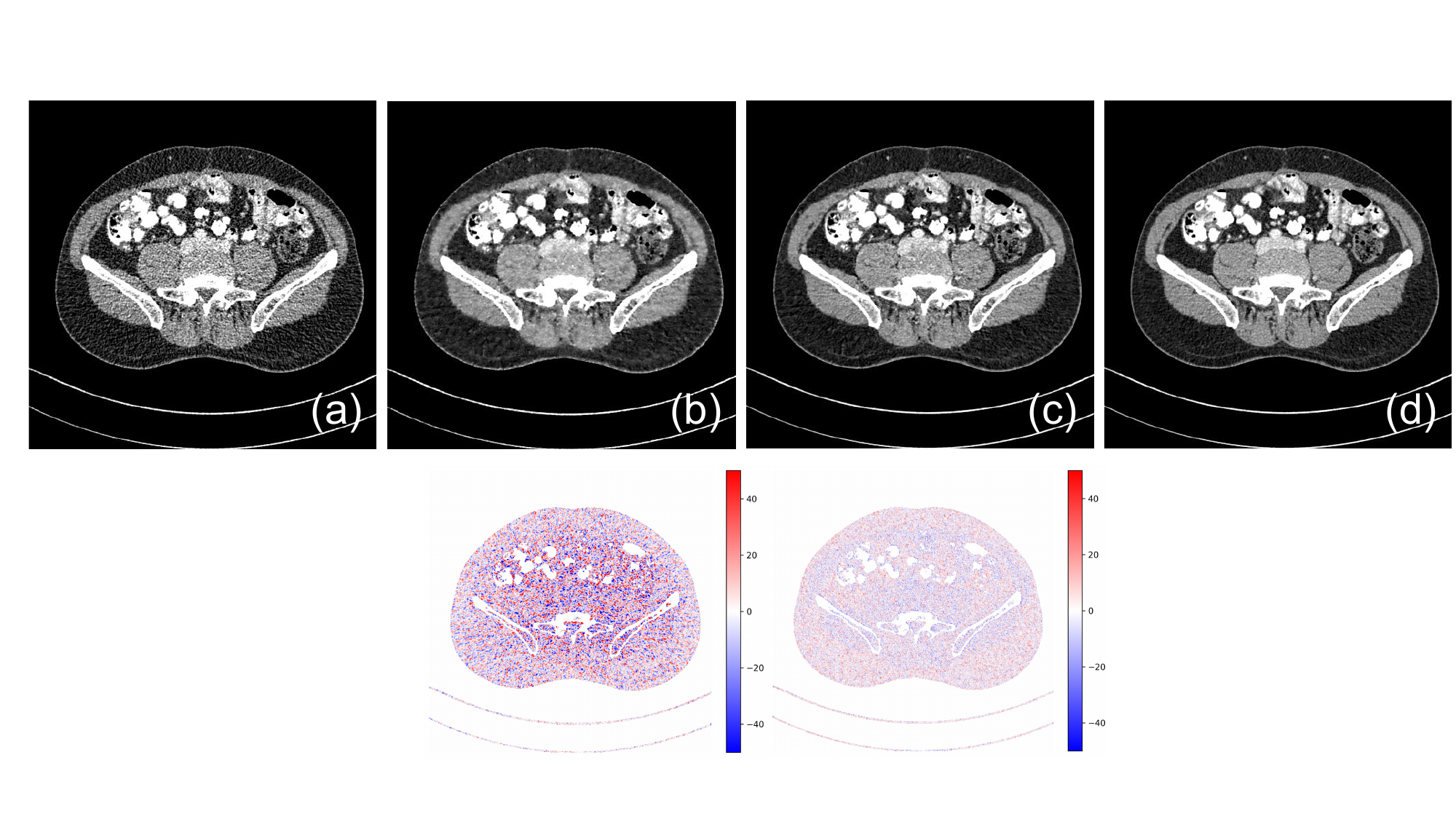}
  \caption{The visualization results of the ablation study. (a) LDCT, (b) w/o Prior, (c) w/ Prior, and (d) NDCT.
  The bottom row shows the absolute error maps between the reconstruction and NDCT, with color intensity indicating the magnitude of the error, where brighter regions correspond to larger errors and darker regions correspond to smaller errors.}
  \Description{The ablation results.}
\label{fig:ablation results}
\end{figure}

\subsection{Ablation Study}
\label{sec:ablation}

We conduct an ablation study to assess the effect of the difference prior. We compare the full model with a baseline conditional diffusion model without the prior. We further ablate the number of reverse steps at inference.

Table~\ref{tab:ablation} and Fig.~\ref{fig:ablation results} show that the difference prior yields the best qualitative and quantitative results. The residual map provides additional constraints during sampling and reduces the tendency to produce spurious structures. We also study the number of reverse steps. The best performance is achieved with 20 steps. This choice provides a good trade-off between reconstruction quality and computational cost. Using fewer or more steps leads to a decline in overall denoising performance.

\section{CONCLUSION}

In this paper, we propose a Residual-Guided Diffusion framework for LDCT denoising. By introducing an explicit difference prior derived from a deterministic coarse estimate, our method effectively constrains diffusion-based denoising process to the residual domain. This approach mitigates the risk of generating spurious textures while preserving fine anatomical details. The coarse-to-fine strategy enables more accurate refinement by leveraging a pre-trained coarse estimate. Extensive experiments across three representative LDCT datasets demonstrate that our method consistently outperforms existing approaches in both quantitative and qualitative assessments. Future work will investigate extending the proposed framework to other tasks, such as super-resolution and deblurring, as well as exploring its applicability to natural images for broader use cases.


\bibliographystyle{ACM-Reference-Format}
\bibliography{references}


\end{document}